\documentclass[aps,pra,reprint,groupedaddress]{revtex4-1}

\usepackage{times}
\usepackage{graphicx}
\usepackage{amssymb}
\usepackage{amsmath}

\newcommand{\qin}{\bar{\bf q}}
\newcommand{\qout}{{\bf q}}
\newcommand{\Tin}{\boldsymbol{\bar{\sf T}}}
\newcommand{\Tout}{\boldsymbol{\sf T}}
\newcommand{\Refl}{\boldsymbol{\sf R}}
\newcommand{\G}{\boldsymbol{\sf G}}
\newcommand{\GE}{\boldsymbol{\cal G}}

\usepackage[colorlinks=true,linkcolor=black,citecolor=blue,urlcolor=blue,hyperfootnotes=true]{hyperref}

\begin{document}


\title{Rectified Lorentz Force from Thermal Current Fluctuations}


\author{Carsten Henkel}
\email[ORCID: 0000-0002-8468-7502, email: ]{carsten.henkel(at)uni-potsdam.de}
\affiliation{Universit\"at Potsdam, Institut f\"ur Physik und Astronomie, Karl-Liebknecht-Str.~24/25,
14476~Potsdam, Germany}


\date{05 February 2024}

\begin{abstract}\noindent
In a conducting medium held at finite temperature, free carriers are performing
Brownian motion and generate fluctuating electromagnetic fields. 
We compute the averaged Lorentz force density that turns out nonzero in a thin
sub-surface layer, pointing towards the surface, while vanishing in the bulk.
This is an elementary example of rectified fluctuations, similar to the Casimir
force or radiative heat transport. 
Our results also provide an experimental way to distinguish between the Drude
and so-called plasma models.
\end{abstract}


\maketitle

\def\OmegaP{\Omega_{p}}
\def\pol{\mu}

\section{Introduction}
\label{s:intro}

The Hall effect is a well-known phenomenon in conducting media 
where a current in a magnetic field generates a transverse voltage
due to the Lorentz force.
Due to the large density of free carriers in conductors, significant
magnetic fields are also internally generated.
The corresponding eddy currents have applications at low frequencies 
for non-invasive material testing (e.g., reduced conductivity at cracks).
Alongside currents induced by oscillating magnetic fields, 
also
the Lorentz 
force plays a role in this context \cite{Uhlig_2012, Li_2018, Alkhalil_2015}.
At frequencies from the infrared through the near-UV, the Lorentz force 
is responsible
for frequency mixing because it is a product of current and field.
This occurs at metal surfaces that provide the necessary
broken symmetry, and leads to, for example, second-harmonic radiation
\cite{Bloembergen_1969, Sipe_1980a, Renger_2009, Renger_2010, Hille_2016}.
A similar phenomenon is optical rectification where typically a short 
and intense laser pulse generates a surge of an electronic current,
providing a source of THz radiation \cite{Kadlec_2004, Hoffmann_2011}.
In samples with inversion symmetry, the electric and magnetic fields of optical pulses may rectify to a quasi-DC electric field that is assisting second-harmonic generation via the third-order Kerr nonlinearity \cite{Trinh_2020}.
Also in these applications, a relatively strong external field
provides the force driving the conduction electrons.

We discuss in this paper the Lorentz (or thermal Hall) force that
arises from the Brownian motion of conduction electrons alone, without
any external perturbation.
A surface is again needed and defines with its normal the distinguished
direction of the fluctuation-averaged (and hence DC) force.
This can be understood as an electromagnetic contribution to the surface
or cleavage energy \cite{Craig_1972, MorgensternHoring_1985, Barton_1986}.
The thermal Hall force will generate some space charge (depletion
zone) below the surface and be balanced by the corresponding electric field.
Experimental indications would therefore be the temperature dependence
of the work function, or a transient change in surface charge density
when the temperature of conduction electrons is pushed up, for example
after absorption of a ultrashort laser 
pulse \cite{Prange_1964,Pudell_2018,Bresson_2020}.

The problem is addressed within the simple setting of fluctuation 
electrodynamics \cite{Rytov_vol3}, and focussing on the local Drude approximation for
the material conductivity. 
The calculations provide an alternative viewpoint on the challenge of defining fluctuation-induced forces inside a macroscopic medium \cite{Griniasty_2017}.
The expression for the averaged Lorentz force contains
two terms one of which would be absent if the so-called plasma model
were used for the metal permittivity.
In line with previous suggestions related to low-frequency magnetic
dipole radiation \cite{Klimchitskaya_2022c,Klimchitskaya_2022b},
the proposed thermal Hall force therefore provides another experimental
clue to understand the anomalous temperature dependence of the Casimir force 
and the unusually large radiative heat transfer on the few-nm scale
\cite{Henkel_2017a,Klimchitskaya_2020a}.

\section{Model}

The electromagnetic force density is given by the familiar expression
\begin{equation}
{\bf f} = \rho {\bf E} + {\bf j} \times {\bf B}
\label{eq:force-density}
\end{equation}
with charge and current densities $\rho$, ${\bf j}$. 
For simplicity, we neglect here pressure terms proportional to the
gradient of the carrier density \cite{Sipe_1980a}
and viscous shear forces \cite{Conti_1999, Hannemann_2021} that lead 
to spatial dispersion (equivalently, a nonlocal conductivity).
If an equilibrium state (with density $e n_0$ and zero current) is perturbed,
the two terms in Eq.\,(\ref{eq:force-density}) are of first and second order, 
respectively, in small deviations from equilibrium.
The Coulomb force leads to the resonance frequency $\Omega_p$ with $\Omega_p^2 = e^2 n_0 / \varepsilon_0 m_e$ for electronic plasma oscillations ($m_e$ is the effective electron mass), 
while the Lorentz force is responsible for second-harmonic generation \cite{Sipe_1980a}.

We consider here the average of the Lorentz force with respect to thermal
fluctuations of charges and fields and derive an integral formula 
for its temperature-dependent DC profile below the surface of a Drude conductor.
The starting point is Rytov's fluctuation electrodynamics \cite{Rytov_vol3} where the electric current density ${\bf j}(x) = {\bf j}({\bf r}, t)$ is a random variable representing both quantum and
thermal fluctuations.
Its symmetrized correlation function is given by the (local) temperature $T$ 
(fluctuation--dissipation theorem)
\begin{align}
  \langle {j}_i(x),\, {j}_k(x') \rangle
  & =
  \tfrac12 \langle {j}_i(x) {j}_k(x') 
                 + {j}_k(x') {j}_i(x) \rangle
  \nonumber\\
  & \hphantom{=
  }
  {} - \langle {j}_i(x) \rangle \, \langle {j}_k(x') \rangle
\label{eq:Rytov-current}
\\
&= \delta_{ik} \delta({\bf r} - {\bf r}')
\int\limits_0^\infty\!\frac{{\rm d}\omega}{2\pi} \, \cos{ \omega(t - t') }
  S_j({\bf r}, \omega)
\nonumber\\
  S_j({\bf r}, \omega) & = 2\hbar\omega \,
  \mathop{\rm Re} \sigma({\bf r}, \omega) \, 
  \coth\frac{\hbar\omega}{2 k_B T} 
\label{eq:Rytov-current-spectrum}
\end{align}
where $\sigma({\bf r}, \omega)$ is the conductivity, assumed local and isotropic.
The Rytov currents generate a magnetic field whose vector potential ${\bf A}$ solves in the transverse gauge the Ampère-Maxwell equation
\begin{equation}
- \nabla^2 {\bf A} - \mu_0 \omega^2 \varepsilon({\bf r}, \omega) {\bf A} = 
\mu_0 {\bf j}_\perp
\label{eq:Ampere-Maxwell}
\end{equation}
with the permittivity $\varepsilon({\bf r}, \omega) = \varepsilon_0 + {\rm i} \sigma({\bf r}, \omega) / \omega$ and the transverse current ${\bf j}_\perp$.
In a homogeneous and isotropic system, 
we expect $\langle {\bf j} \times {\bf B} \rangle = {\bf 0}$, 
since there is no preferred direction
(see also Ref.\,\cite{Griniasty_2017}).
We therefore focus in the following on a simple half-space geometry 
with the metal filling $z \ge 0$. 
Parallel to the surface, a Fourier expansion with wave vector ${\bf Q} = (q_x, q_y)$ is applied where rotational invariance around the surface normal may be assumed. 
At fixed ${\bf Q}$, the vector potential is given by a Green tensor
\begin{align}
{\bf A}({\bf Q}, z) & = \int_0^\infty\!{\rm d}z' \, 
\G({\bf Q}, z, z') \cdot {\bf j}({\bf Q}, z')
\label{eq:Green-function-A-j-1}
\\
\G({\bf Q}, z, z') &= \frac{ {\rm i} \mu_0 }{ 2 q }
\big(
   \Tin \, {\rm e}^{ - {\rm i} q z }
+  \Refl \Tin \, {\rm e}^{ + {\rm i} q z }
\big) \, {\rm e}^{ {\rm i} q z' }
\quad \text{for $z < z'$}
\nonumber\\
 &= \frac{ {\rm i} \mu_0 }{ 2 q }
\big(
   \Tout \, {\rm e}^{ - {\rm i} q z' }
+  \Refl \Tin \, {\rm e}^{+ {\rm i} q z' }
\big) \, {\rm e}^{ {\rm i} q z }
\quad \text{for $z' < z$}
\label{eq:Green-function-A-j}
\end{align}
where $q^2 = \mu_0 \omega^2 \varepsilon(\omega) - Q^2$. 
This $q$ with $\mathop{\rm Re} q, \mathop{\rm Im} q \ge 0$ provides the normal component of the wave vectors $\qout = {\bf Q} + q {\bf e}_z$, $\qin = {\bf Q} - q {\bf e}_z$ for reflected and incident waves, respectively.
The tensors $\Tout$, $\Tin$ are projectors transverse to $\qout$, $\qin$.
The tensor $\Refl$ describes the fields reflected from the inner surface.
It is diagonal when expanded into principal transverse polarisations (p/TM and s/TE),
and contains the reflection amplitudes $r_p$, $r_s$.
The average of the vector product ${\bf j} \times {\bf B}$ 
  with respect to
the Rytov currents gives with the local and isotropic correlation~(\ref{eq:Rytov-current}) a vector structure proportional to
\begin{equation}
\langle {\bf j}^* \times [\qout \times (\Tin\, {\bf j})] \rangle \propto
\mathop{\rm tr}( \Tin ) \, \qout - \Tin \, \qout
\label{eq:average-cross-product}
\end{equation}
with analogous formulas for $\qin$, $\Refl\Tin$ etc. 
If the tensor $\Tout$ corres\-ponds to $\qout$, the last term vanishes by transversality.
After the integral over the in-plane angle of ${\bf Q}$, only components normal to the
surface remain.

Working through the polarisation vectors (see Appendix~\ref{a:pol-vectors} 
for details), we indeed find that the fluctuation-averaged 
Lorentz force density $\langle {\bf j} \times {\bf B} \rangle = f \, {\bf e}_z$ is 
orthogonal to the surface and given by
\begin{equation}
f = 
- \frac{ \mu_0 }{ 4\pi }
\int\limits_0^\infty\!{\rm d}\omega\, S_j(\omega) 
\mathop{\rm Re}
\int\limits_0^\infty\!Q\,{\rm d}Q\,{\rm e}^{ 2 i q z } ( r_p + r_s )
\label{eq:integral-jxB-1}
\end{equation}
The current spectrum $S_j$ is given in Eq.\,(\ref{eq:Rytov-current-spectrum}).
We are going to use the Drude model for the conductivity
\begin{equation}
\sigma(\omega) = \frac{ \sigma_0 }{ 1 - {\rm i} \omega \tau }
\label{eq:Drude-conductivity}
\end{equation}
with the DC conductivity $\sigma_0$ and the scattering (collision) rate $1/\tau$.
  This model describes well any conducting material between DC and below additional resonance frequencies. The latter may correspond to optically active phonons (typically in the infrared) or interband transitions (in the visible and above) and depend on the material \cite{DresselGruener}.
The so-called plasma model corresponds to 
  the limit 
$\sigma_0, \tau \to \infty$ at fixed 
  plasma frequency
$\Omega_p^2 = \sigma_0 / (\varepsilon_0 \tau)$.
Physical realisations of this model are superconducting materials below their gap frequency and at temperatures much below critical.
Its characteristic feature is a purely imaginary conductivity, except at zero frequency. 
The weight of the corresponding $\delta$-function,
\begin{equation}
\mathop{\rm Re}\sigma(\omega)
=
\frac{ \sigma_0/\tau^2 }{ 1/\tau^2 + \omega^2 }
\to \pi \, \varepsilon_0 \Omega_p^2 \, \delta(\omega)
\label{eq:delta-omega-conductivity}
\end{equation}
has been attributed to the density of superconducting carriers (Cooper pairs) \cite{Berlinsky_1993}, and is generally temperature-dependent.

The reflection coefficients from the ``inner'' side of a metal-vacuum interface are 
in the Fresnel approximation
\begin{align}
r_p &= \frac{ \varepsilon v - \varepsilon_0 q }{ \varepsilon v + \varepsilon_0 q }
\nonumber
\\
r_s &= \frac{ q - v }{ q + v }
\,,\qquad
v = \sqrt{ (\omega/c)^2 - Q^2 }
\label{eq:Fresnel-formulas}
\end{align}
where $\varepsilon_0$ is the vacuum permittivity.

The calculation above focussed on the contribution from fluctuating currents.
Within fluctuation electrodynamics, another contribution arises from
fluctuating fields \cite{Rytov_vol3}.
To provide a simple motivation for this additional term, consider a toy model with just two normal mode amplitudes $a, b$. 
By construction, these are uncorrelated.
Two generic fields $A, B$ can be written as linear combination of normal modes, 
$A = c_1 a + c_2 b$ and $B = d_1 a + d_2 b$.
They have a correlation function
\begin{equation}
\langle A^* B \rangle = 
  c_1^* d_1 \, \langle a^* a \rangle
+ c_2^* d_2 \, \langle b^* b \rangle
\label{eq:}
\end{equation}
To connect the coefficients in this expression with measurable quantities, we attribute the term $c_1 a = A_{\rm fl}$ to ``fluctuations'' and $c_2 b = A_{\rm ind}$ to an ``induced'' field, and similarly $d_1 a = B_{\rm ind}$ and $d_2 b = B_{\rm fl}$.
Such an identification appears naturally when equations of motion are linearised around an equilibrium situation, in particular in the context of Langevin equations.
With these notations, the correlation becomes
\begin{align}
\langle A^* B \rangle & = 
  \frac{ d_1 }{ c_1 } \, \langle A_{\rm fl}^* A_{\rm fl} \rangle
+ \frac{ c_2^* }{ d_2^* } \, \langle B_{\rm fl}^* B_{\rm fl} \rangle
\nonumber\\
& =
  \frac{ \partial B_{\rm ind} }{ \partial A_{\rm fl} } \, \langle A_{\rm fl}^* A_{\rm fl} \rangle
+ \frac{ \partial A_{\rm ind}^* }{ \partial B_{\rm fl}^* } \, \langle B_{\rm fl}^* B_{\rm fl} \rangle
\label{eq:Rytov-split-and-correlation}
\end{align}
In the last step, we have expressed the ratio $d_1 / c_1$ by the linear response of variable $B$ to $A$ and \emph{vice versa}.
With respect to the calculation performed so far, the term $\langle A_{\rm fl}^* A_{\rm fl} \rangle$ in Eq.\,(\ref{eq:Rytov-split-and-correlation}) corresponds to current fluctuations, 
and $\partial B_{\rm ind} /\partial A_{\rm fl}$ describes the magnetic field generated by them. 
The second term $\langle B_{\rm fl}^* B_{\rm fl} \rangle$ corresponds to magnetic field fluctuations that we now turn to.

The current responds to ${\bf B}_{\rm fl}$ via the associated electric field and 
Ohm's law  
${\bf j}^{\rm ind} = \sigma \, {\bf E}^{\rm fl}$.
The thermal Lorentz force is thus
determined by the average Poynting vector 
$\langle {\bf E}^{\rm fl} \times {\bf B}^{\rm fl} \rangle$.
We express the spectrum of field fluctuations with the fluctuation--dissipation
theorem, assuming thermal equilibrium at temperature $T$. 
For our purposes, this temperature coincides with the electron temperature
because the field responds very quickly to its sources, in virtue of its wide
continuous mode spectrum.
Working through the corresponding calculations (Appendix \ref{a:field-fluctuations}), 
we find that an expression 
similar to Eq.\,(\ref{eq:integral-jxB-1}) has to be added to the Lorentz force. 
The full result has the explicit form
\begin{align}
\mbox{total:}\quad
f(z, T) & =
- \frac{ \hbar \mu_0 }{ 2\pi }
\mathop{\rm Re}
\int\limits_0^\infty\!{\rm d}\omega\, 
\bigg[
\omega \, 
\sigma(\omega) \,
\coth\frac{\hbar\omega}{2 k_B T} \, \times
\nonumber\\
& \qquad
\times \int\limits_0^\infty\!Q\,{\rm d}Q\,{\rm e}^{ 2 i q z } ( r_p + r_s )
\bigg]
\label{eq:integral-jxB}
\end{align}
This is the main result of the present paper. We discuss its properties in the following.

\section{Discussion}

\subsection{General features}

\begin{figure}[htbp] 
   \centering
   \includegraphics[width=\columnwidth]{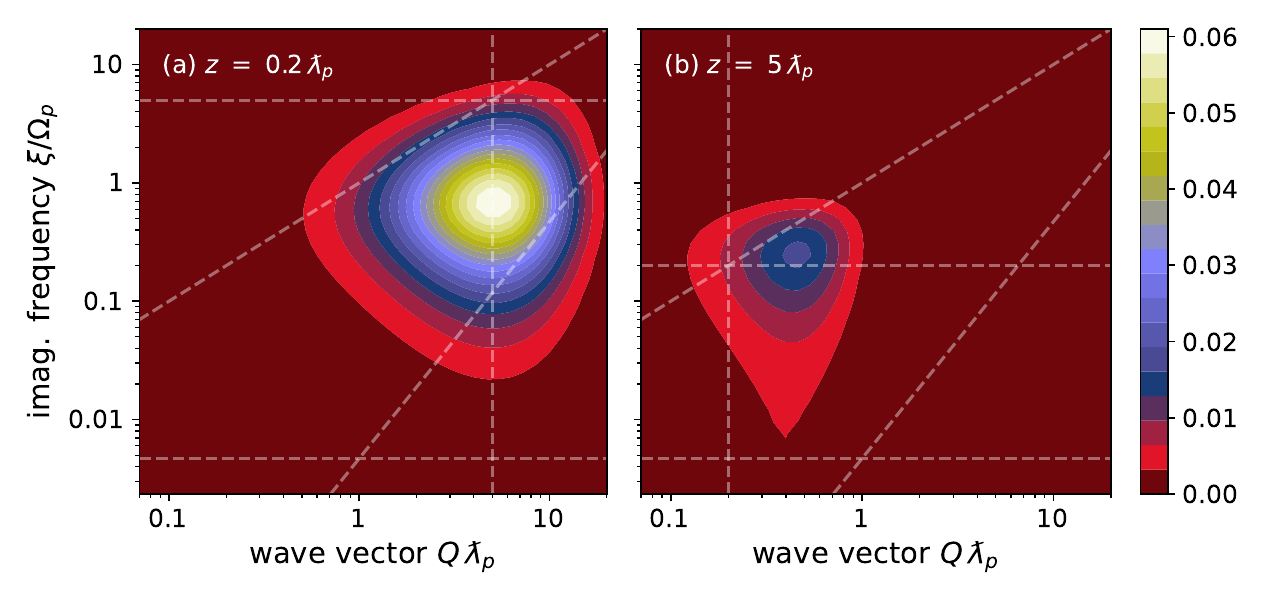} 
   \caption[]{%
Integrand of the average Lorentz force due to quantum fluctuations ($T = 0$, arbitrary units). 
A Wick rotation to imaginary frequencies $\xi$ has been applied. 
Left (a) and right (b): short and large distance as indicated.
Parameters: 
plasma frequency $\Omega_p \approx 210/\tau$ (typical for Au)
and wavelength $\lambdabar_p = c/\Omega_p$.
The dashed lines in (a, b) mark the values 
$\xi = c/z$, 
$\xi = c Q$ (light cone), 
$\xi = Q^2 / (\mu_0\sigma_0)$ (magnetic diffusion),
$\xi = 1/\tau$, 
and $Q = 1/z$.
To reduce the dynamics of the data points, the integrand has been multiplied by $z^3$.
}
   \label{fig:integrand-quantum}
\end{figure}

A net force appears only due to the reflection from the surface at $z = 0$, 
as expected from broken rotational symmetry.
Similar to the Casimir effect, the Lorentz force contains a pure quantum contribution
that is ultraviolet dominated, since $\coth\frac12\beta\omega \to 1$ at high frequencies. 
In practice, the UV transparency of the material makes this contribution finite.
Indeed, from the sum of the two Fresnel coefficients
\begin{equation}
r_p + r_s = \frac{ 2 v q ( \varepsilon - \varepsilon_0 ) }{ (\varepsilon v + q) (q + v) }
\label{eq:}
\end{equation}
it appears explicitly that the integrand decays sufficiently fast at high 
frequencies. 
This is illustrated in Figs.\,\ref{fig:integrand-quantum} and~\ref{fig:spectral_map}
where the integrand of Eq.\,(\ref{eq:integral-jxB}) is plotted. 

In the zero-temperature limit, it is expedient to shift the frequency integration to the imaginary axis, 
$\omega = {\rm i} \xi$.
In this representation, large frequencies and wave vectors are exponentially damped by the factor ${\rm e}^{ 2 {\rm i} q z } \approx
\exp[ - 2 (z/c)\sqrt{\Omega_p^2 + \xi^2 + c^2 Q^2}]$.
(This approximation assumes $\xi \gg 1/\tau$.)
A rough estimation of the double integral yields a scaling of the average Lorentz force density according to
\begin{equation}
T = 0: \quad
f(z, 0) \sim \frac{ \hbar \Omega_p }{ \lambdabar_p \, z^3 }
\label{eq:quantum-Lorentz}
\end{equation}
We expect both the plasma and the Drude model to give comparable contributions, unless distances larger than $c \tau \gg \lambdabar_p$ are considered.
In addition, for frequencies in the visible range and above, it is mandatory to take into account deviations from the Drude (or plasma) models, using, e.g., tabulated optical data \cite{Klimchitskaya_2009}.
A more detailed discussion is left to future work.

Deep in the bulk, $z \to + \infty$, the exponential ${\rm e}^{2{\rm i} q z}$ makes the force vanish.
Since the medium wave vector $q$ in Eq.\,(\ref{eq:integral-jxB}) is complex, we may expect an oscillatory behavior.
The exponential ${\rm e}^{2i q z}$ becomes approximately real 
deeply below the light cone ($Q \gg \omega / c$).
The typical long-range behaviour in the infrared is $q \approx (1+i)/\delta$ with the skin depth $\delta^2(\omega) = 2 / (\mu_0 \sigma_0 \omega)$. 
This corresponds to the diffusive 
propagation
of magnetic fields in a conducting medium.

The limit $z \to 0$ is beyond the local (Drude or plasma) model 
because $r_p$ tends towards a constant at large $Q$, spoiling convergence. 
This is cured when using a nonlocal (${\bf q}$-dependent) conductivity 
whose magnitude drops for short-wavelength fields. 
The leading-order behaviour in the local approximation is discussed below.

\subsection{Thermal Hall force}

\begin{figure}[tbp]
   \centering
   \raisebox{0mm}{\includegraphics[height=0.2\textwidth]{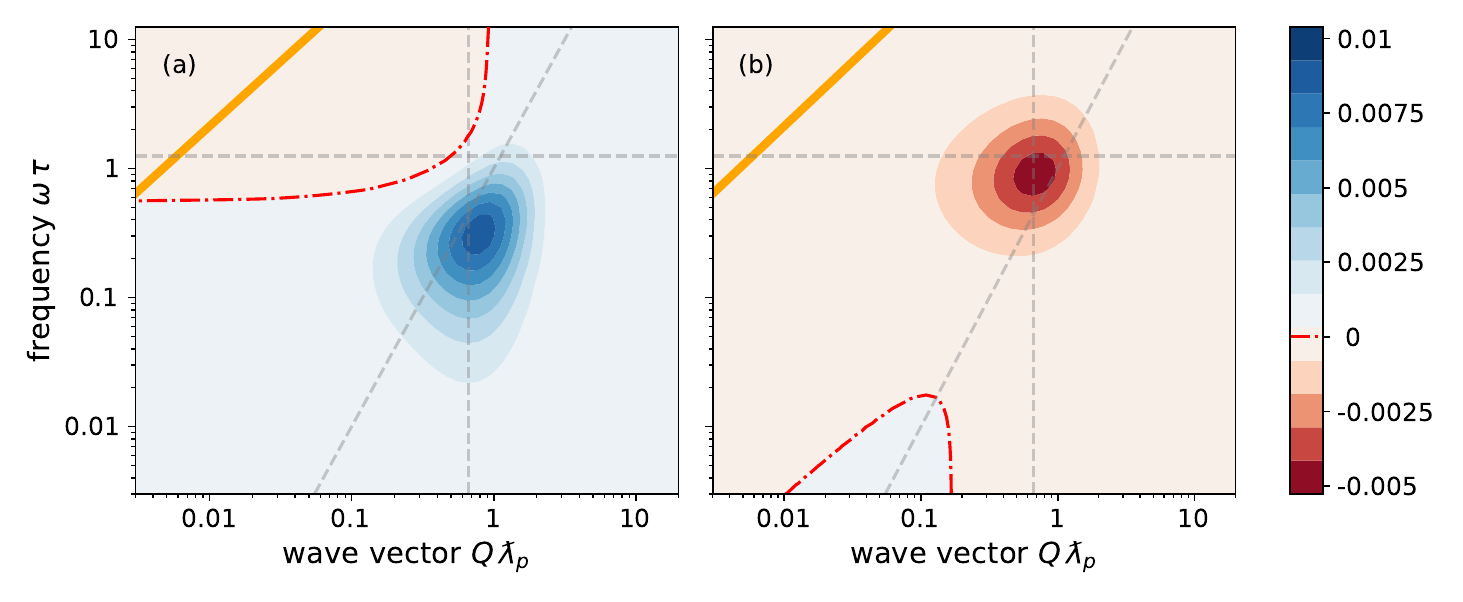}}
   \hspace*{3mm}
   \includegraphics*[height=0.2\textwidth]{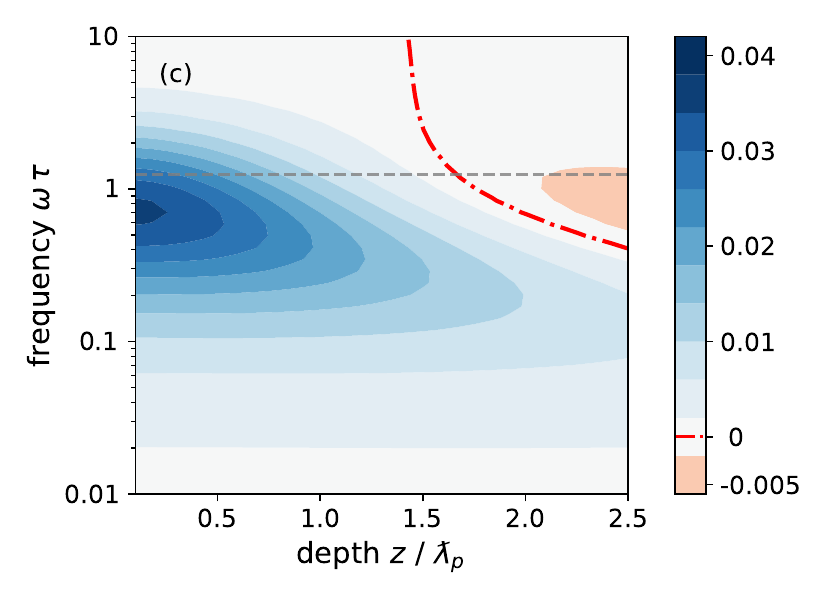} 
\caption[]{Spectrum of the thermal Lorentz force density (arbitrary units, 
real frequencies).
Top row (a, b): integrand of Eq.\,(\ref{eq:integral-jxB}),
with the $T = 0$ contribution subtracted; 
in panel (b), only the imaginary part of the conductivity is kept (similar to the 
plasma model).
Bottom (c): spectrum $f(\omega, z)$ before evaluating the $\omega$-integral.
  Sign changes occur at the red dash-dotted lines.
Parameters: 
temperature $k_B T = 1.25\,\hbar/\tau$,
plasma frequency $\Omega_p \approx 210/\tau$ (typical for Au),
distance $z = 1.5\,\lambdabar_p$ in (a, b).
The dashed lines in (a, b) mark the values $Q = 1/z$, $\hbar \omega = k_B T$,
$\omega = \mu_0 \sigma_0 Q^2$ (magnetic diffusivity), 
in solid orange
the light cone $\omega = c Q$.
To reduce the dynamics of the data points in (c), the force has been multiplied by $z^2$.
}
\label{fig:spectral_map}
\end{figure}

In the following, we subtract 
the quantum contribution, $\coth(\hbar \omega / 2 k_B T) - 1 = 2 \bar{n}(\omega/T)$,
so that 
the thermal component of the Lorentz force 
is 
proportional to the Bose-Einstein distribution $\bar{n}(\omega/T)$.
It is dominated by frequencies with $\hbar\omega \lesssim k_B T$ (mid infrared and below, 
see Fig.\,\ref{fig:spectral_map}(c)).
The plots in Fig.\,\ref{fig:spectral_map}(a, b) illustrate that the
integrand of Eq.\,(\ref{eq:integral-jxB}) in the $(Q,\omega)$-plane (panel (a))
would change sign if only the term due to field fluctuations were kept (panel (b)).

Note that in the plasma model, where the conductivity is purely imaginary, the
integrand is nonzero only above the light cone ($\omega > c Q$) and approximately above 
the plasma frequency $\Omega_p$. 
Otherwise, the medium wave vector $q$ is purely imaginary, and the reflection
coefficients $r_s, r_p$ turn out real.
This severely suppresses the thermal contribution to the average Lorentz force,
since for typical temperatures, we have $\hbar \Omega_p \gg k_B T$.
It is therefore instructive to evaluate the contribution from the singular DC conductivity of Eq.\,(\ref{eq:delta-omega-conductivity}).
In calculations along imaginary frequencies, using a generalised plasma model, this term generates a permittivity $\varepsilon({\rm i}\xi) \sim \Omega_p^2/\xi^2$, either by inserting Eq.\,(\ref{eq:delta-omega-conductivity}) into Kramers-Kronig relations or, more carefully, by first isolating the zero-frequency pole \cite{Klimchitskaya_2007c, Levine_2008}.
A physical interpretation in terms of current fluctuations for superconductors is not obvious, however.
Fields penetrate into a superconducting medium down to roughly the same depth (the plasma wavelength $\lambdabar_p$) as the layer where the thermal Lorentz force is nonzero, see Fig.\,\ref{fig:scaled-force} below.
But one would expect from the Meißner effect that in the bulk of a sample, there are neither static currents nor magnetic fields. 
In Ref.\,\cite{Henkel_2009a}, Intravaia and the present author suggested to interpret the fluctuation electrodynamics of a medium with Eq.\,(\ref{eq:delta-omega-conductivity}) in terms of an ``ideal conductor'' model.
Its bulk is filled with ``frozen currents'' and concomitant magnetic field loops.
Inserting the conductivity~(\ref{eq:delta-omega-conductivity}) into Eq.\,(\ref{eq:integral-jxB-1}), we get for the thermal Lorentz force the expression
\begin{align}
\mbox{ideal cond.:} \quad
\Delta f(z, T) = &- 
%
\frac{k_B T}{\lambdabar_p^2}
\int\limits_0^\infty\!{\rm d}Q\,{\rm e}^{ -2 Q z } 
\frac{ Q \, \kappa }{ \kappa + Q }
\nonumber\\
& \hphantom{
}
+ \mbox{exp. small terms}
\label{eq:ideal-conductor-force}
\end{align}
with 
the plasma wavelength $\lambdabar_p = c/\Omega_p$ and
$\kappa^2 = (\Omega_p/c)^2 + Q^2$.
The integral here has the asymptotic form $1/(8z^2)$ [$1/(4z^2)$]
for $z \ll \lambdabar_p$ [$z \gg \lambdabar_p$], respectively,
the same scaling as the Coulomb force due to image charges.
The exponentially small terms arise from frequencies $\hbar\omega \gtrsim \hbar\Omega_p \gg k_B T$. 
The resulting force is shown in dash-dotted 
in Fig.\,\ref{fig:scaled-force} below.

\begin{figure}[htbp]
   \centering
   \includegraphics*[width=0.4\textwidth]{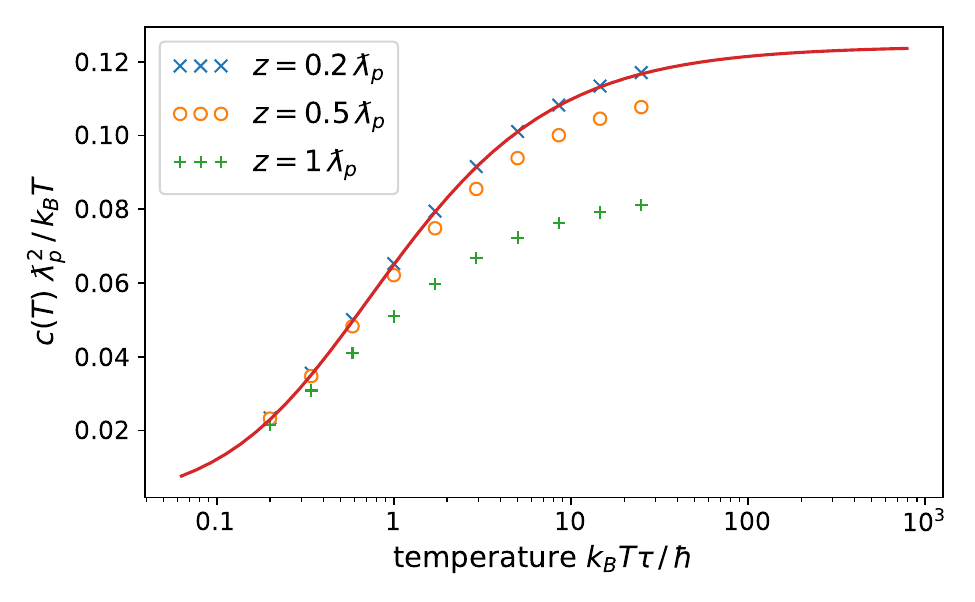} 
\caption[]{Temperature dependence of the amplitude $c(T)$ of the rectified Lorentz force density $f \approx - c(T)/z^2$ at short distances, normalised to $k_B T / \lambda_p^2$. 
Solid line: Eq.\,(\ref{eq:approximation-small-z}), 
symbols: numerical integration of Eq.\,(\ref{eq:integral-jxB}) with the $T = 0$ contribution subtracted.
Material parameters as in Fig.\,\ref{fig:spectral_map}.
(Note that $\tau$ is not temperature-dependent here.)}
\label{fig:prefactor-c(T)}
\end{figure}

In good metallic conductors, the reflection coefficients are dominated by $|r_p| \approx 1$ while 
$r_s \approx - \tfrac14(\varepsilon - 1) (\omega/c Q)^2 \to 0$
for large $Q \gg 
|\varepsilon| \omega / c,\,
\omega / c$ (evanescent waves).
This allows for an approximate evaluation of the $Q$-integral in Eq.\,(\ref{eq:integral-jxB}).
  We drop $r_s$ in the leading order and get again
the scaling law $f \sim -1/z^2$, the same as the Coulomb force due to image charges.
We have checked that this captures well the short-distance behaviour of the force density, 
$f(z, T) \approx -c(T) / z^2$,
with a prefactor given by
\begin{align}
c(T) & \approx 
\frac{ \hbar \mu_0 \sigma_0 }{ 4\pi }
\int\limits_0^\infty\!{\rm d}\omega\, \frac{ \omega\, \bar{n}(\omega/T) }{ 1 + \omega^2 \tau^2 }
\nonumber\\
&= 
\frac{ k_B T }{ 8\pi \lambdabar_p^2 }
\left(
\beta \log \frac{\beta}{2\pi} - \pi - \beta \, \psi(\beta/2\pi)
\right)
\label{eq:approximation-small-z}
\end{align}
Here, $\beta = \hbar / (k_B T \, \tau)$ and $\psi(\cdot)$ is the digamma function.
Recall that $\tau$ is the scattering time in the Drude conductivity, 
and $\bar{n}(\omega/T)$ the Bose-Einstein distribution.
This expression is shown in Fig.\,\ref{fig:prefactor-c(T)} after dividing out the scale factors $k_B T / \lambdabar_p^2$: we observe only a minor dynamics, 
even though the product $k_B T \tau / \hbar$ varies over three orders of magnitude.
The agreement with the full numerical integration is excellent at the short distance $z = 0.2\,\lambdabar_p$.

The distance dependence at fixed temperature can be read off from Fig.\,\ref{fig:scaled-force} where the combination $-f(z, T) \, z^2 / (k_B T)$ is shown.
The force decays into the bulk with strongly damped oscillations, of which remains only a crossing of the curves for different temperatures at a depth $z \approx 3.5\,\lambdabar_p$.
Beyond this depth, the linear scaling with temperature becomes exact. 
The rectified Lorentz force is thus restricted to a few plasma penetration depths, typically about $100\,{\rm nm}$.
The ideal conductor also gives a scaling linear in $T$, but the weak modifications relative to the $1/z^2$ power law display the opposite trend.

\begin{figure}[htbp]
   \centering
   \includegraphics*[width=0.37\textwidth]{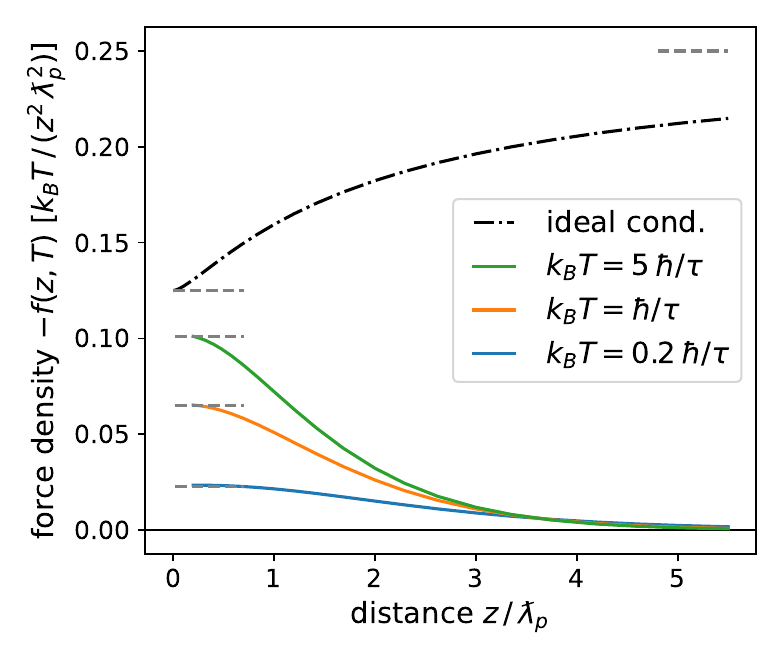} 
\caption[]{Distance dependence of the DC force density, normalised to $T / z^2$ 
and with flipped sign, for different temperatures. 
Black dash-dotted line: ideal conductor result [Eq.\,(\ref{eq:ideal-conductor-force})]. 
Coloured lines: Drude conductor with finite damping time $\tau$.
The dashed gray lines give the 
  short- and large-distance limits quoted after Eq.\,(\ref{eq:ideal-conductor-force}) and 
the short-distance limit of Eq.\,(\ref{eq:approximation-small-z}).
Same parameters as in Fig.\,\ref{fig:spectral_map},
they correspond for typical conductors like Au to
$\hbar/\tau \approx 400\,{\rm K}$ and $\lambdabar_p = c/\Omega_p \approx 20\,{\rm nm}$.}
\label{fig:scaled-force}
\end{figure}

\subsection{Physical consequences}

Among the physical consequences suggested by this prediction, 
we mentioned in the Introduction a temperature-dependent shift $\Delta \phi(T)$ 
in the work function of a metal. 
Indeed, the Lorentz force is pulling charges towards the surface.
To calculate the corresponding energy gain, 
we need to regularise the $1/z^2$ divergence as $z \to 0$. 
This is physically achieved by adopting a non-local dielectric function (spatial
dispersion), as discussed in Refs.\,\cite{DresselGruener, Sernelius_2005a, Svetovoy06a}.
A characteristic length scale related to the compressibility of the electron gas
is the Debye screening length $\ell_D = v_F / \Omega_p$ where $v_F$ is typically of the 
order of the Fermi velocity.

If we integrate the Lorentz force density 
  from $z = \infty$ down
to a cutoff at $z = \ell_D$ 
and divide by the equilibrium carrier density $n_0$, we get the following estimate 
\begin{equation}
\Delta \phi(T) \approx - \frac{ c(T) }{ n_0 \ell_D } \approx
- 0.06 \, k_B T \frac{e^2}{\varepsilon_0 \hbar c} \frac{ \hbar/\lambdabar_p }{ m v_F } 
\label{eq:estimate-change-in-work-function}
\end{equation}
Both fractions on the rhs are smaller than unity:
  the first is $4\pi/137 \approx 0.0917$,
  and for Gold, the second takes the value $\approx 0.00380$.
But a Kelvin probe locked to a
periodic temperature modulation may prove sufficiently sensitive.

A complementary phenomenon is the induced sub-surface space charge that screens the thermal Lorentz force, restoring electro-chemical equilibrium.
From the Coulomb law, its cumulative density $\Delta Q/A$ per unit area is of the order of
\begin{equation}
\frac{ \Delta Q }{ A } \approx 
\frac{ \varepsilon_0 }{ e n_0 } \lim_{z \to 0} f(z) \approx
- 0.06 \frac{ e }{ \lambdabar_p^2 } \frac{ k_B T }{ m v_F^2 }
\label{eq:sub-surface-screening-charge}
\end{equation}
This is again a quite small charge, barely an elementary charge per square micron for Gold.
If this charge shows fluctuations in the MHz frequency band, however, these may be detectable with miniaturised ion traps because the corresponding fluctuations in the Coulomb force work against laser cooling the ion to its motional ground state \cite{Brownnutt_2015}.

\section{Conclusion}

We have explored in this paper a thermal Hall effect arising from the correlation
between current density and magnetic field in a conducting medium at finite temperature.
It turns out that in a thin layer below the material surface (its thickness being
comparable to the Meissner penetration depth $\lambdabar_p$), the Lorentz force
density, averaged over thermal fluctuations, is nonzero and points towards 
the surface, similar to the interaction with image charges.
We found that a Drude model gives a distinct prediction compared to the so-called 
plasma model because the corresponding force spectra have opposite signs 
[see Fig.\,\ref{fig:spectral_map}(a,b)].
The thermal Hall voltage is relatively small, however. 

The next step could be the regularisation on short length scales, using a spatially
dispersive permittivity and suitable boundary conditions.
Another interesting perspective is the fluctuation spectrum of the Lorentz 
force around its thermal average, that arises from fourth-order correlations of Rytov 
currents. 
This may provide an alternative, physical picture for the unusual electric field fluctuations
observed in ion traps (anomalous heating) that are often attributed to surface
contaminations \cite{Brownnutt_2015}.

\begin{acknowledgments}
This research was supported in part by the National Science Foundation under Grant No.\ PHY-1748958. 
I thank the KITP (University of California at Santa Barbara) for its hospitality while
attending the 2022 summer program ``Quantum and Thermal Electrodynamic Fluctuations in the Presence of Matter: Progress and Challenges''.
I acknowledge funding by the Deutsche Forschungsgemeinschaft (DFG, German Research Foundation) within CRC/SFB 1636 ``Elementary processes of light-driven reactions at nano-scale metals'' (Project ID 510943930, Project No. A01).
\end{acknowledgments}

\appendix


\section{Details of the Calculation}
\label{a:calculation}

\subsection{Polarisation vectors}
\label{a:pol-vectors}

The following transverse polarisation vectors are used to expand 
the transverse projection tensor $\Tout = {\bf e}_s \otimes {\bf e}_s + 
{\bf e}_p \otimes {\bf e}_p$
\begin{equation}
{\bf e}_s = \hat{\bf Q} \times {\bf e}_z
\qquad
{\bf e}_p = ( q \hat{\bf Q} - Q {\bf e}_z ) / k 
\label{eq:def-pol-vectors}
\end{equation}
where $\hat{\bf Q}$ is the unit vector parallel to ${\bf Q}$
and $k = \omega [\mu_0 \varepsilon(\omega)]^{1/2}$.
For the wave vector $\qin$ of the incident wave (orthogonal projector $\Tin$), 
we use the mirror images
\begin{equation}
\bar{\bf e}_s = {\bf e}_s
\qquad
\bar{\bf e}_p = (q \hat{\bf Q} + Q {\bf e}_z) / k
\label{eq:def-pol-vectors-flip}
\end{equation}
This leads to the following compact form of the transverse reflection
tensor \cite{Sipe_1987}
\begin{equation}
\Refl \Tin = r_s \, {\bf e}_s \otimes \bar{\bf e}_s + 
r_p \, {\bf e}_p \otimes \bar{\bf e}_p
\label{eq:}
\end{equation}
As a consistency check, consider the limit of normal incidence where
both polarisations behave in the same way.

According to Eq.\,(\ref{eq:average-cross-product}), we need the trace 
of this tensor
\begin{equation}
\mathop{\rm tr} \Refl \Tin = r_s + r_p \, (q^2 - Q^2) / k^2
\label{eq:}
\end{equation}
and the image of the reflected wave vector
\begin{equation}
\Refl \Tin \cdot \qout = 2 r_p \, {\bf e}_p \, q Q / k
\label{eq:}
\end{equation}
This is nonzero because $\qin$ and $\qout$ differ by one mirror reflection 
from the surface.
We proceed to the angular integration over the in-plane angle $\varphi$ 
of ${\bf Q}$.
The reflection coefficients only depend on its magnitude $Q$.
We have
\begin{equation}
\int\!\frac{{\rm d}\varphi}{2\pi} \qout = q \, {\bf e}_z
\qquad
\int\!\frac{{\rm d}\varphi}{2\pi} {\bf e}_p = - (Q/k) {\bf e}_z
\label{eq:phi-averages}
\end{equation}
so that after integrating over $\varphi$, Eq.\,(\ref{eq:average-cross-product}) becomes
\begin{align}
&\int\!\frac{{\rm d}\varphi}{2\pi} \left[
\mathop{\rm tr}( \Refl \Tin ) \, \qout - \Refl \Tin \cdot \qout
\right] 
\nonumber\\
& = 
  q \left[ r_s + r_p \, (q^2 - Q^2) / k^2 \right] \, {\bf e}_z
+ 2 q \, r_p \, (Q^2 / k^2) \, {\bf e}_z
\nonumber\\
& = q \left( r_s + r_p \right) {\bf e}_z
\label{eq:}
\end{align}
We still have to multiply with the phase factor ${\rm e}^{ 2 i q z }$ 
from the Green function~(\ref{eq:Green-function-A-j}). 
  The terms without the reflection coefficients (homogeneous medium) cancel thanks to the first integral in Eq.\,(\ref{eq:phi-averages}):
  we combine the limits $z' \searrow z$ and $z' \nearrow z$ and exploit the local current correlation function~(\ref{eq:Rytov-current}) to evaluate the  $z'$-integral.
Taking into account the symmetrised correlation function, eventually 
introduces a real part \cite{Agarwal_1975a}, and we get Eq.\,(\ref{eq:integral-jxB-1}).

\subsection{Average Poynting vector}
\label{a:field-fluctuations}

As outlined after Eq.\,(\ref{eq:Fresnel-formulas}), the contribution of
field rather than current fluctuations involves the calculation of the
correlation function
$\langle {\bf E}^*( {\bf r}, \omega ) \times {\bf B}( {\bf r}, \omega' ) \rangle$.
Using the Faraday equation to express the magnetic field, we have to evaluate
\begin{align}
& \langle {\bf E}^*( {\bf r}, \omega ) \times [
\nabla' \times {\bf E}( {\bf r}', \omega' ) ] \rangle
\label{eq:Poynting-and-EE}
\\
& =
\nabla'
\langle {\bf E}^*( {\bf r}, \omega ) \cdot {\bf E}( {\bf r}', \omega' ) \rangle
-
\langle [{\bf E}^*( {\bf r}, \omega ) \cdot \nabla' ] \, 
{\bf E}( {\bf r}', \omega' ) \rangle
\nonumber
\end{align}
eventually taking the limit \mbox{${\bf r}' \to {\bf r}$}. 
The electric field autocorrelation is given by the fluc\-tua\-tion-dissipation 
theorem~\cite{Rytov_vol3, Landau_vol9, Buhmann_vol2}
\begin{equation}
\langle E^*_i( {\bf r}, \omega ) E_j( {\bf r}', \omega' ) \rangle
= \frac{ 4\pi \hbar \, \delta(\omega - \omega') 
}{ {\rm e}^{ \hbar\omega / k_BT} - 1 }
\mathop{\rm Im} {\cal G}_{ij}( {\bf r}, {\bf r}', \omega )
\label{eq:FD-theorem-EE}
\end{equation}
We assume here for simplicity the medium to be reciprocal so that
${\cal G}_{ij}( {\bf r}, {\bf r}', \omega ) = {\cal G}_{ji}( {\bf r}', {\bf r}, \omega )$.
Recall that this Green tensor determines the electric field ${\bf E}({\bf r}, \omega)$
radiated by a monochromatic point dipole of amplitude ${\bf d}$ located at 
position ${\bf r}'$ in the medium, ${\bf E} = \GE \cdot {\bf d}$.

The Green tensor splits into a part relevant for a homogeneous bulk 
medium that only depends on the difference ${\bf r} - {\bf r}'$. 
Its derivatives vanish for ${\bf r}' \to {\bf r}$.
The remaining part near a planar surface can be written with reflection
coefficients (Weyl expansion, $z, z' \ge 0$) \cite{Sipe_1987}
\begin{equation}
\GE^{\rm refl}( {\bf r}, {\bf r}', \omega ) = 
{\rm i} \mu_0 \omega^2
\int\!\frac{ {\rm d}^2Q }{ (2\pi)^2 }
\frac{ 
    {\rm e}^{ {\rm i} (\qout \cdot {\bf r} - \qin \cdot {\bf r}') }
}{ 2 q }
\Refl \Tin 
\label{eq:}
\end{equation}
Performing the derivatives of Eq.\,(\ref{eq:Poynting-and-EE}) under the
imaginary part of this expression, leads to a quite similar calculation
as in Sec.\,\ref{a:pol-vectors} and results in
\begin{align}
& \nabla' \mathop{\rm Im}
\mathop{\rm tr} \GE^{\rm refl}( {\bf r}, {\bf r}', \omega )
-
\sum_{i,j} \frac{\partial}{ \partial x_i' } \mathop{\rm Im}
{\cal G}_{ij}^{\rm refl}( {\bf r}, {\bf r}', \omega )\,
{\bf e}_j 
\nonumber\\[-2ex]
& =
- \frac{\mu_0}{4\pi} {\bf e}_z \, \omega^2 \mathop{\rm Im}
\int\limits_{0}^{\infty} \! {\rm d}Q\, Q \, 
{\rm e}^{ 2 {\rm i} q z } \left( r_s + r_p \right)
\label{eq:}
\end{align}
as ${\bf r}' \to {\bf r}$.
The final steps are to multiply this with $- i \sigma^* / \omega$  to
convert ${\bf E}^*$ into ${\bf j}^*$ and $\nabla \times {\bf E}$ into
${\bf B}$ [see Eq.\,(\ref{eq:Poynting-and-EE})], 
and to take the real part to get the symmetrised correlation.
This makes the imaginary part of the conductivity appear. 
Writing the frequency integral over positive frequencies only, leads
in conjunction with Eq.\,(\ref{eq:integral-jxB-1})
to the final result~(\ref{eq:integral-jxB}).


\begin{thebibliography}{38}%
\makeatletter
\providecommand \@ifxundefined [1]{%
 \@ifx{#1\undefined}
}%
\providecommand \@ifnum [1]{%
 \ifnum #1\expandafter \@firstoftwo
 \else \expandafter \@secondoftwo
 \fi
}%
\providecommand \@ifx [1]{%
 \ifx #1\expandafter \@firstoftwo
 \else \expandafter \@secondoftwo
 \fi
}%
\providecommand \natexlab [1]{#1}%
\providecommand \enquote  [1]{``#1''}%
\providecommand \bibnamefont  [1]{#1}%
\providecommand \bibfnamefont [1]{#1}%
\providecommand \citenamefont [1]{#1}%
\providecommand \href@noop [0]{\@secondoftwo}%
\providecommand \href [0]{\begingroup \@sanitize@url \@href}%
\providecommand \@href[1]{\@@startlink{#1}\@@href}%
\providecommand \@@href[1]{\endgroup#1\@@endlink}%
\providecommand \@sanitize@url [0]{\catcode `\\12\catcode `\$12\catcode
  `\&12\catcode `\#12\catcode `\^12\catcode `\_12\catcode `\%12\relax}%
\providecommand \@@startlink[1]{}%
\providecommand \@@endlink[0]{}%
\providecommand \url  [0]{\begingroup\@sanitize@url \@url }%
\providecommand \@url [1]{\endgroup\@href {#1}{\urlprefix }}%
\providecommand \urlprefix  [0]{URL }%
\providecommand \Eprint [0]{\href }%
\providecommand \doibase [0]{http://dx.doi.org/}%
\providecommand \selectlanguage [0]{\@gobble}%
\providecommand \bibinfo  [0]{\@secondoftwo}%
\providecommand \bibfield  [0]{\@secondoftwo}%
\providecommand \translation [1]{[#1]}%
\providecommand \BibitemOpen [0]{}%
\providecommand \bibitemStop [0]{}%
\providecommand \bibitemNoStop [0]{.\EOS\space}%
\providecommand \EOS [0]{\spacefactor3000\relax}%
\providecommand \BibitemShut  [1]{\csname bibitem#1\endcsname}%
\let\auto@bib@innerbib\@empty
\bibitem [{\citenamefont {Uhlig}\ \emph {et~al.}(2012)\citenamefont {Uhlig},
  \citenamefont {Zec}, \citenamefont {Brauer}, and\ \citenamefont
  {Thess}}]{Uhlig_2012}%
  \BibitemOpen
  \bibfield  {author} {\bibinfo {author} {\bibfnamefont {R.~P.}\ \bibnamefont
  {Uhlig}}, \bibinfo {author} {\bibfnamefont {M.}~\bibnamefont {Zec}}, \bibinfo
  {author} {\bibfnamefont {H.}~\bibnamefont {Brauer}},  and\ \bibinfo {author}
  {\bibfnamefont {A.}~\bibnamefont {Thess}},\ }\href {\doibase
  10.1007/s10921-012-0147-7} {\bibfield  {journal} {\bibinfo  {journal} {J.
  Nondestruct. Eval.}\ }\textbf {\bibinfo {volume} {31}},\ \bibinfo {pages}
  {357} (\bibinfo {year} {2012})}\BibitemShut {NoStop}%
\bibitem [{\citenamefont {Li}\ \emph {et~al.}(2018)\citenamefont {Li},
  \citenamefont {Liu}, \citenamefont {Dai}, \citenamefont {Zhang}, and\
  \citenamefont {Wang}}]{Li_2018}%
  \BibitemOpen
  \bibfield  {author} {\bibinfo {author} {\bibfnamefont {C.}~\bibnamefont
  {Li}}, \bibinfo {author} {\bibfnamefont {R.}~\bibnamefont {Liu}}, \bibinfo
  {author} {\bibfnamefont {S.}~\bibnamefont {Dai}}, \bibinfo {author}
  {\bibfnamefont {N.}~\bibnamefont {Zhang}},  and\ \bibinfo {author}
  {\bibfnamefont {X.}~\bibnamefont {Wang}},\ }\href {\doibase
  10.3390/app8112289} {\bibfield  {journal} {\bibinfo  {journal} {Appl. Sci.}\
  }\textbf {\bibinfo {volume} {8}},\ \bibinfo {pages} {2289} (\bibinfo {year}
  {2018})}\BibitemShut {NoStop}%
\bibitem [{\citenamefont {Alkhalil}\ \emph {et~al.}(2015)\citenamefont
  {Alkhalil}, \citenamefont {Kolesnikov}, and\ \citenamefont
  {Thess}}]{Alkhalil_2015}%
  \BibitemOpen
  \bibfield  {author} {\bibinfo {author} {\bibfnamefont {S.}~\bibnamefont
  {Alkhalil}}, \bibinfo {author} {\bibfnamefont {Y.}~\bibnamefont
  {Kolesnikov}},  and\ \bibinfo {author} {\bibfnamefont {A.}~\bibnamefont
  {Thess}},\ }\href {\doibase 10.1088/0957-0233/26/11/115605} {\bibfield
  {journal} {\bibinfo  {journal} {Meas. Sci. Technol.}\ }\textbf {\bibinfo
  {volume} {26}},\ \bibinfo {pages} {115605} (\bibinfo {year}
  {2015})}\BibitemShut {NoStop}%
\bibitem [{\citenamefont {Bloembergen}\ \emph {et~al.}(1969)\citenamefont
  {Bloembergen}, \citenamefont {Chang}, \citenamefont {Jha}, and\
  \citenamefont {Lee}}]{Bloembergen_1969}%
  \BibitemOpen
  \bibfield  {author} {\bibinfo {author} {\bibfnamefont {N.}~\bibnamefont
  {Bloembergen}}, \bibinfo {author} {\bibfnamefont {R.~K.}\ \bibnamefont
  {Chang}}, \bibinfo {author} {\bibfnamefont {S.~S.}\ \bibnamefont {Jha}}, \
  and\ \bibinfo {author} {\bibfnamefont {C.~H.}\ \bibnamefont {Lee}},\ }\href
  {\doibase 10.1103/PhysRev.178.1528.3} {\bibfield  {journal} {\bibinfo
  {journal} {Phys. Rev.}\ }\textbf {\bibinfo {volume} {178}},\ \bibinfo {pages}
  {1528} (\bibinfo {year} {1969})}\BibitemShut {NoStop}%
\bibitem [{\citenamefont {Sipe}\ \emph {et~al.}(1980)\citenamefont {Sipe},
  \citenamefont {So}, \citenamefont {Fukui}, and\ \citenamefont
  {Stegeman}}]{Sipe_1980a}%
  \BibitemOpen
  \bibfield  {author} {\bibinfo {author} {\bibfnamefont {J.~E.}\ \bibnamefont
  {Sipe}}, \bibinfo {author} {\bibfnamefont {V.~C.~Y.}\ \bibnamefont {So}},
  \bibinfo {author} {\bibfnamefont {M.}~\bibnamefont {Fukui}},  and\ \bibinfo
  {author} {\bibfnamefont {G.~I.}\ \bibnamefont {Stegeman}},\ }\href {\doibase
  10.1103/PhysRevB.21.4389} {\bibfield  {journal} {\bibinfo  {journal} {Phys.
  Rev. B}\ }\textbf {\bibinfo {volume} {21}},\ \bibinfo {pages} {4389}
  (\bibinfo {year} {1980})}\BibitemShut {NoStop}%
\bibitem [{\citenamefont {Renger}\ \emph {et~al.}(2009)\citenamefont {Renger},
  \citenamefont {Quidant}, \citenamefont {van Hulst}, \citenamefont {Palomba},\
  and\ \citenamefont {Novotny}}]{Renger_2009}%
  \BibitemOpen
  \bibfield  {author} {\bibinfo {author} {\bibfnamefont {J.}~\bibnamefont
  {Renger}}, \bibinfo {author} {\bibfnamefont {R.}~\bibnamefont {Quidant}},
  \bibinfo {author} {\bibfnamefont {N.}~\bibnamefont {van Hulst}}, \bibinfo
  {author} {\bibfnamefont {S.}~\bibnamefont {Palomba}},  and\ \bibinfo
  {author} {\bibfnamefont {L.}~\bibnamefont {Novotny}},\ }\href {\doibase
  10.1103/PhysRevLett.103.266802} {\bibfield  {journal} {\bibinfo  {journal}
  {Phys. Rev. Lett.}\ }\textbf {\bibinfo {volume} {103}},\ \bibinfo {pages}
  {266802} (\bibinfo {year} {2009})}\BibitemShut {NoStop}%
\bibitem [{\citenamefont {Renger}\ \emph {et~al.}(2010)\citenamefont {Renger},
  \citenamefont {Quidant}, \citenamefont {van Hulst}, and\ \citenamefont
  {Novotny}}]{Renger_2010}%
  \BibitemOpen
  \bibfield  {author} {\bibinfo {author} {\bibfnamefont {J.}~\bibnamefont
  {Renger}}, \bibinfo {author} {\bibfnamefont {R.}~\bibnamefont {Quidant}},
  \bibinfo {author} {\bibfnamefont {N.}~\bibnamefont {van Hulst}},  and\
  \bibinfo {author} {\bibfnamefont {L.}~\bibnamefont {Novotny}},\ }\href
  {\doibase 10.1103/PhysRevLett.104.046803} {\bibfield  {journal} {\bibinfo
  {journal} {Phys. Rev. Lett.}\ }\textbf {\bibinfo {volume} {104}},\ \bibinfo
  {pages} {046803} (\bibinfo {year} {2010})}\BibitemShut {NoStop}%
\bibitem [{\citenamefont {Hille}\ \emph {et~al.}(2016)\citenamefont {Hille},
  \citenamefont {Moeferdt}, \citenamefont {Wolff}, \citenamefont {Matyssek},
  \citenamefont {Rodr{\'\i}guez-Oliveros}, \citenamefont {Prohm}, \citenamefont
  {Niegemann}, \citenamefont {Grafstr{\"o}m}, \citenamefont {Eng}, and\
  \citenamefont {Busch}}]{Hille_2016}%
  \BibitemOpen
  \bibfield  {author} {\bibinfo {author} {\bibfnamefont {A.}~\bibnamefont
  {Hille}}, \bibinfo {author} {\bibfnamefont {M.}~\bibnamefont {Moeferdt}},
  \bibinfo {author} {\bibfnamefont {C.}~\bibnamefont {Wolff}}, \bibinfo
  {author} {\bibfnamefont {C.}~\bibnamefont {Matyssek}}, \bibinfo {author}
  {\bibfnamefont {R.}~\bibnamefont {Rodr{\'\i}guez-Oliveros}}, \bibinfo
  {author} {\bibfnamefont {C.}~\bibnamefont {Prohm}}, \bibinfo {author}
  {\bibfnamefont {J.}~\bibnamefont {Niegemann}}, \bibinfo {author}
  {\bibfnamefont {S.}~\bibnamefont {Grafstr{\"o}m}}, \bibinfo {author}
  {\bibfnamefont {L.~M.}\ \bibnamefont {Eng}},  and\ \bibinfo {author}
  {\bibfnamefont {K.}~\bibnamefont {Busch}},\ }\href {\doibase
  10.1021/acs.jpcc.5b08348} {\bibfield  {journal} {\bibinfo  {journal} {J.
  Phys. Chem. C}\ }\textbf {\bibinfo {volume} {120}},\ \bibinfo {pages} {1163}
  (\bibinfo {year} {2016})}\BibitemShut {NoStop}%
\bibitem [{\citenamefont {Kadlec}\ \emph {et~al.}(2004)\citenamefont {Kadlec},
  \citenamefont {Ku{\v z}el}, and\ \citenamefont {Coutaz}}]{Kadlec_2004}%
  \BibitemOpen
  \bibfield  {author} {\bibinfo {author} {\bibfnamefont {F.}~\bibnamefont
  {Kadlec}}, \bibinfo {author} {\bibfnamefont {P.}~\bibnamefont {Ku{\v z}el}},
   and\ \bibinfo {author} {\bibfnamefont {J.-L.}\ \bibnamefont {Coutaz}},\
  }\href {\doibase 10.1364/ol.29.002674} {\bibfield  {journal} {\bibinfo
  {journal} {Opt. Lett.}\ }\textbf {\bibinfo {volume} {29}},\ \bibinfo {pages}
  {2674} (\bibinfo {year} {2004})}\BibitemShut {NoStop}%
\bibitem [{\citenamefont {Hoffmann} and\ \citenamefont
  {F{\"u}l{\"o}p}(2011)}]{Hoffmann_2011}%
  \BibitemOpen
  \bibfield  {author} {\bibinfo {author} {\bibfnamefont {M.~C.}\ \bibnamefont
  {Hoffmann}} and\ \bibinfo {author} {\bibfnamefont {J.~A.}\ \bibnamefont
  {F{\"u}l{\"o}p}},\ }\href {\doibase 10.1088/0022-3727/44/8/083001} {\bibfield
   {journal} {\bibinfo  {journal} {J. Phys. D: Appl. Phys.}\ }\textbf {\bibinfo
  {volume} {44}},\ \bibinfo {pages} {083001} (\bibinfo {year}
  {2011})}\BibitemShut {NoStop}%
\bibitem [{\citenamefont {Trinh}\ \emph {et~al.}(2020)\citenamefont {Trinh},
  \citenamefont {Smail}, \citenamefont {Makhal}, \citenamefont {Yang},
  \citenamefont {Kim}, and\ \citenamefont {Rand}}]{Trinh_2020}%
  \BibitemOpen
  \bibfield  {author} {\bibinfo {author} {\bibfnamefont {M.~T.}\ \bibnamefont
  {Trinh}}, \bibinfo {author} {\bibfnamefont {G.}~\bibnamefont {Smail}},
  \bibinfo {author} {\bibfnamefont {K.}~\bibnamefont {Makhal}}, \bibinfo
  {author} {\bibfnamefont {D.~S.}\ \bibnamefont {Yang}}, \bibinfo {author}
  {\bibfnamefont {J.}~\bibnamefont {Kim}},  and\ \bibinfo {author}
  {\bibfnamefont {S.~C.}\ \bibnamefont {Rand}},\ }\href {\doibase
  10.1038/s41467-020-19125-w} {\bibfield  {journal} {\bibinfo  {journal}
  {Nature Commun.}\ }\textbf {\bibinfo {volume} {11}},\ \bibinfo {pages} {5296}
  (\bibinfo {year} {2020})}\BibitemShut {NoStop}%
\bibitem [{\citenamefont {Craig}(1972)}]{Craig_1972}%
  \BibitemOpen
  \bibfield  {author} {\bibinfo {author} {\bibfnamefont {R.~A.}\ \bibnamefont
  {Craig}},\ }\href {\doibase 10.1103/PhysRevB.6.1134} {\bibfield  {journal}
  {\bibinfo  {journal} {Phys. Rev. B}\ }\textbf {\bibinfo {volume} {6}},\
  \bibinfo {pages} {1134} (\bibinfo {year} {1972})}\BibitemShut {NoStop}%
\bibitem [{\citenamefont {Morgenstern~Horing}\ \emph
  {et~al.}(1985)\citenamefont {Morgenstern~Horing}, \citenamefont {Kamen},\
  and\ \citenamefont {Gumbs}}]{MorgensternHoring_1985}%
  \BibitemOpen
  \bibfield  {author} {\bibinfo {author} {\bibfnamefont {N.~J.}\ \bibnamefont
  {Morgenstern~Horing}}, \bibinfo {author} {\bibfnamefont {E.}~\bibnamefont
  {Kamen}},  and\ \bibinfo {author} {\bibfnamefont {G.}~\bibnamefont
  {Gumbs}},\ }\href {\doibase 10.1103/PhysRevB.31.8269} {\bibfield  {journal}
  {\bibinfo  {journal} {Phys. Rev. B}\ }\textbf {\bibinfo {volume} {31}},\
  \bibinfo {pages} {8269} (\bibinfo {year} {1985})}\BibitemShut {NoStop}%
\bibitem [{\citenamefont {Barton}(1986)}]{Barton_1986}%
  \BibitemOpen
  \bibfield  {author} {\bibinfo {author} {\bibfnamefont {G.}~\bibnamefont
  {Barton}},\ }\href {\doibase 10.1088/0022-3719/19/7/009} {\bibfield
  {journal} {\bibinfo  {journal} {J. Phys. C: Sol. State Phys.}\ }\textbf
  {\bibinfo {volume} {19}},\ \bibinfo {pages} {975} (\bibinfo {year}
  {1986})}\BibitemShut {NoStop}%
\bibitem [{\citenamefont {Prange} and\ \citenamefont
  {Kadanoff}(1964)}]{Prange_1964}%
  \BibitemOpen
  \bibfield  {author} {\bibinfo {author} {\bibfnamefont {R.~E.}\ \bibnamefont
  {Prange}} and\ \bibinfo {author} {\bibfnamefont {L.~P.}\ \bibnamefont
  {Kadanoff}},\ }\href {\doibase 10.1103/PhysRev.134.A566} {\bibfield
  {journal} {\bibinfo  {journal} {Phys. Rev.}\ }\textbf {\bibinfo {volume}
  {134}},\ \bibinfo {pages} {A566} (\bibinfo {year} {1964})}\BibitemShut
  {NoStop}%
\bibitem [{\citenamefont {Pudell}\ \emph {et~al.}(2018)\citenamefont {Pudell},
  \citenamefont {Maznev}, \citenamefont {Herzog}, \citenamefont {Kronseder},
  \citenamefont {Back}, \citenamefont {Malinowski}, \citenamefont {von
  Reppert}, and\ \citenamefont {Bargheer}}]{Pudell_2018}%
  \BibitemOpen
  \bibfield  {author} {\bibinfo {author} {\bibfnamefont {J.}~\bibnamefont
  {Pudell}}, \bibinfo {author} {\bibfnamefont {A.}~\bibnamefont {Maznev}},
  \bibinfo {author} {\bibfnamefont {M.}~\bibnamefont {Herzog}}, \bibinfo
  {author} {\bibfnamefont {M.}~\bibnamefont {Kronseder}}, \bibinfo {author}
  {\bibfnamefont {C.}~\bibnamefont {Back}}, \bibinfo {author} {\bibfnamefont
  {G.}~\bibnamefont {Malinowski}}, \bibinfo {author} {\bibfnamefont
  {A.}~\bibnamefont {von Reppert}},  and\ \bibinfo {author} {\bibfnamefont
  {M.}~\bibnamefont {Bargheer}},\ }\href {\doibase 10.1038/s41467-018-05693-5}
  {\bibfield  {journal} {\bibinfo  {journal} {Nature Commun.}\ }\textbf
  {\bibinfo {volume} {9}},\ \bibinfo {pages} {3335} (\bibinfo {year}
  {2018})}\BibitemShut {NoStop}%
\bibitem [{\citenamefont {Bresson}\ \emph {et~al.}(2020)\citenamefont
  {Bresson}, \citenamefont {Bryche}, \citenamefont {Besbes}, \citenamefont
  {Moreau}, \citenamefont {Karsenti}, \citenamefont {Charette}, \citenamefont
  {Morris}, and\ \citenamefont {Canva}}]{Bresson_2020}%
  \BibitemOpen
  \bibfield  {author} {\bibinfo {author} {\bibfnamefont {P.}~\bibnamefont
  {Bresson}}, \bibinfo {author} {\bibfnamefont {J.-F.}\ \bibnamefont {Bryche}},
  \bibinfo {author} {\bibfnamefont {M.}~\bibnamefont {Besbes}}, \bibinfo
  {author} {\bibfnamefont {J.}~\bibnamefont {Moreau}}, \bibinfo {author}
  {\bibfnamefont {P.-L.}\ \bibnamefont {Karsenti}}, \bibinfo {author}
  {\bibfnamefont {P.~G.}\ \bibnamefont {Charette}}, \bibinfo {author}
  {\bibfnamefont {D.}~\bibnamefont {Morris}},  and\ \bibinfo {author}
  {\bibfnamefont {M.}~\bibnamefont {Canva}},\ }\href {\doibase
  10.1103/PhysRevB.102.155127} {\bibfield  {journal} {\bibinfo  {journal}
  {Phys. Rev. B}\ }\textbf {\bibinfo {volume} {102}},\ \bibinfo {pages}
  {155127} (\bibinfo {year} {2020})}\BibitemShut {NoStop}%
\bibitem [{\citenamefont {Rytov}\ \emph {et~al.}(1989)\citenamefont {Rytov},
  \citenamefont {Kravtsov}, and\ \citenamefont {Tatarskii}}]{Rytov_vol3}%
  \BibitemOpen
  \bibfield  {author} {\bibinfo {author} {\bibfnamefont {S.~M.}\ \bibnamefont
  {Rytov}}, \bibinfo {author} {\bibfnamefont {Y.~A.}\ \bibnamefont {Kravtsov}},
   and\ \bibinfo {author} {\bibfnamefont {V.~I.}\ \bibnamefont {Tatarskii}},\
  }\href@noop {} {\emph {\bibinfo {title} {Elements of Random Fields}}},\
  \bibinfo {series} {Principles of Statistical Radiophysics}, Vol.~\bibinfo
  {volume} {3}\ (\bibinfo  {publisher} {Springer},\ \bibinfo {address}
  {Berlin},\ \bibinfo {year} {1989})\BibitemShut {NoStop}%
\bibitem [{\citenamefont {Griniasty} and\ \citenamefont
  {Leonhardt}(2017)}]{Griniasty_2017}%
  \BibitemOpen
  \bibfield  {author} {\bibinfo {author} {\bibfnamefont {I.}~\bibnamefont
  {Griniasty}} and\ \bibinfo {author} {\bibfnamefont {U.}~\bibnamefont
  {Leonhardt}},\ }\href {\doibase 10.1103/PhysRevA.96.032123} {\bibfield
  {journal} {\bibinfo  {journal} {Phys. Rev. A}\ }\textbf {\bibinfo {volume}
  {96}},\ \bibinfo {pages} {032123} (\bibinfo {year} {2017})}\BibitemShut
  {NoStop}%
\bibitem [{\citenamefont {Klimchitskaya}\ \emph
  {et~al.}(2022{\natexlab{a}})\citenamefont {Klimchitskaya}, \citenamefont
  {Mostepanenko}, and\ \citenamefont {Svetovoy}}]{Klimchitskaya_2022c}%
  \BibitemOpen
  \bibfield  {author} {\bibinfo {author} {\bibfnamefont {G.~L.}\ \bibnamefont
  {Klimchitskaya}}, \bibinfo {author} {\bibfnamefont {V.~M.}\ \bibnamefont
  {Mostepanenko}},  and\ \bibinfo {author} {\bibfnamefont {V.~B.}\
  \bibnamefont {Svetovoy}},\ }\href {\doibase 10.3390/universe8110574}
  {\bibfield  {journal} {\bibinfo  {journal} {Universe}\ }\textbf {\bibinfo
  {volume} {8}},\ \bibinfo {pages} {574} (\bibinfo {year}
  {2022}{\natexlab{a}})}\BibitemShut {NoStop}%
\bibitem [{\citenamefont {Klimchitskaya}\ \emph
  {et~al.}(2022{\natexlab{b}})\citenamefont {Klimchitskaya}, \citenamefont
  {Mostepanenko}, and\ \citenamefont {Svetovoy}}]{Klimchitskaya_2022b}%
  \BibitemOpen
  \bibfield  {author} {\bibinfo {author} {\bibfnamefont {G.~L.}\ \bibnamefont
  {Klimchitskaya}}, \bibinfo {author} {\bibfnamefont {V.~M.}\ \bibnamefont
  {Mostepanenko}},  and\ \bibinfo {author} {\bibfnamefont {V.~B.}\
  \bibnamefont {Svetovoy}},\ }\href {\doibase 10.1209/0295-5075/ac8c69}
  {\bibfield  {journal} {\bibinfo  {journal} {Europhys. Lett.}\ }\textbf
  {\bibinfo {volume} {139}},\ \bibinfo {pages} {66001} (\bibinfo {year}
  {2022}{\natexlab{b}})}\BibitemShut {NoStop}%
\bibitem [{\citenamefont {Henkel}(2017)}]{Henkel_2017a}%
  \BibitemOpen
  \bibfield  {author} {\bibinfo {author} {\bibfnamefont {C.}~\bibnamefont
  {Henkel}},\ }\href {\doibase 10.1515/zna-2016-0372} {\bibfield  {journal}
  {\bibinfo  {journal} {Z. f. Naturforsch. A}\ }\textbf {\bibinfo {volume}
  {72}},\ \bibinfo {pages} {99} (\bibinfo {year} {2017})}\BibitemShut {NoStop}%
\bibitem [{\citenamefont {Klimchitskaya} and\ \citenamefont
  {Mostepanenko}(2020)}]{Klimchitskaya_2020a}%
  \BibitemOpen
  \bibfield  {author} {\bibinfo {author} {\bibfnamefont {G.~L.}\ \bibnamefont
  {Klimchitskaya}} and\ \bibinfo {author} {\bibfnamefont {V.~M.}\ \bibnamefont
  {Mostepanenko}},\ }\href {\doibase 10.1142/S0217732320400076} {\bibfield
  {journal} {\bibinfo  {journal} {Mod. Phys. Lett. A}\ }\textbf {\bibinfo
  {volume} {35}},\ \bibinfo {pages} {2040007} (\bibinfo {year}
  {2020})}\BibitemShut {NoStop}%
\bibitem [{\citenamefont {Conti} and\ \citenamefont
  {Vignale}(1999)}]{Conti_1999}%
  \BibitemOpen
  \bibfield  {author} {\bibinfo {author} {\bibfnamefont {S.}~\bibnamefont
  {Conti}} and\ \bibinfo {author} {\bibfnamefont {G.}~\bibnamefont
  {Vignale}},\ }\href {\doibase 10.1103/PhysRevB.60.7966} {\bibfield  {journal}
  {\bibinfo  {journal} {Phys. Rev. B}\ }\textbf {\bibinfo {volume} {60}},\
  \bibinfo {pages} {7966} (\bibinfo {year} {1999})}\BibitemShut {NoStop}%
\bibitem [{\citenamefont {Hannemann}\ \emph {et~al.}(2021)\citenamefont
  {Hannemann}, \citenamefont {Wegner}, and\ \citenamefont
  {Henkel}}]{Hannemann_2021}%
  \BibitemOpen
  \bibfield  {author} {\bibinfo {author} {\bibfnamefont {M.}~\bibnamefont
  {Hannemann}}, \bibinfo {author} {\bibfnamefont {G.}~\bibnamefont {Wegner}}, \
  and\ \bibinfo {author} {\bibfnamefont {C.}~\bibnamefont {Henkel}},\ }\href
  {\doibase 10.3390/universe7040108} {\bibfield  {journal} {\bibinfo  {journal}
  {Universe}\ }\textbf {\bibinfo {volume} {7}},\ \bibinfo {pages} {108}
  (\bibinfo {year} {2021})},\ \Eprint {http://arxiv.org/abs/arXiv:2104.00334}
  {arXiv:2104.00334} \BibitemShut {NoStop}%
\bibitem [{\citenamefont {Dressel} and\ \citenamefont
  {Gr{\"u}ner}(2002)}]{DresselGruener}%
  \BibitemOpen
  \bibfield  {author} {\bibinfo {author} {\bibfnamefont {M.}~\bibnamefont
  {Dressel}} and\ \bibinfo {author} {\bibfnamefont {G.}~\bibnamefont
  {Gr{\"u}ner}},\ }\href {\doibase 10.1017/CBO9780511606168} {\emph {\bibinfo
  {title} {{E}lectrodynamics of {S}olids -- {O}ptical {P}roperties of
  {E}lectrons in {M}atter}}}\ (\bibinfo  {publisher} {Cambridge University
  Press},\ \bibinfo {address} {Cambridge},\ \bibinfo {year} {2002})\BibitemShut
  {NoStop}%
\bibitem [{\citenamefont {Berlinsky}\ \emph {et~al.}(1993)\citenamefont
  {Berlinsky}, \citenamefont {Kallin}, \citenamefont {Rose}, and\
  \citenamefont {Shi}}]{Berlinsky_1993}%
  \BibitemOpen
  \bibfield  {author} {\bibinfo {author} {\bibfnamefont {A.~J.}\ \bibnamefont
  {Berlinsky}}, \bibinfo {author} {\bibfnamefont {C.}~\bibnamefont {Kallin}},
  \bibinfo {author} {\bibfnamefont {G.}~\bibnamefont {Rose}},  and\ \bibinfo
  {author} {\bibfnamefont {A.-C.}\ \bibnamefont {Shi}},\ }\href {\doibase
  10.1103/PhysRevB.48.4074} {\bibfield  {journal} {\bibinfo  {journal} {Phys.
  Rev. B}\ }\textbf {\bibinfo {volume} {48}},\ \bibinfo {pages} {4074}
  (\bibinfo {year} {1993})}\BibitemShut {NoStop}%
\bibitem [{\citenamefont {Klimchitskaya}\ \emph {et~al.}(2009)\citenamefont
  {Klimchitskaya}, \citenamefont {Mohideen}, and\ \citenamefont
  {Mostepanenko}}]{Klimchitskaya_2009}%
  \BibitemOpen
  \bibfield  {author} {\bibinfo {author} {\bibfnamefont {G.~L.}\ \bibnamefont
  {Klimchitskaya}}, \bibinfo {author} {\bibfnamefont {U.}~\bibnamefont
  {Mohideen}},  and\ \bibinfo {author} {\bibfnamefont {V.~M.}\ \bibnamefont
  {Mostepanenko}},\ }\href {\doibase 10.1103/RevModPhys.81.1827} {\bibfield
  {journal} {\bibinfo  {journal} {Rev. Mod. Phys.}\ }\textbf {\bibinfo {volume}
  {81}},\ \bibinfo {pages} {1827} (\bibinfo {year} {2009})}\BibitemShut
  {NoStop}%
\bibitem [{\citenamefont {Klimchitskaya}\ \emph {et~al.}(2007)\citenamefont
  {Klimchitskaya}, \citenamefont {Mohideen}, and\ \citenamefont
  {Mostepanenko}}]{Klimchitskaya_2007c}%
  \BibitemOpen
  \bibfield  {author} {\bibinfo {author} {\bibfnamefont {G.~L.}\ \bibnamefont
  {Klimchitskaya}}, \bibinfo {author} {\bibfnamefont {U.}~\bibnamefont
  {Mohideen}},  and\ \bibinfo {author} {\bibfnamefont {V.~M.}\ \bibnamefont
  {Mostepanenko}},\ }\href {\doibase 10.1088/1751-8113/40/17/F04} {\bibfield
  {journal} {\bibinfo  {journal} {J. Phys. A}\ }\textbf {\bibinfo {volume}
  {40}},\ \bibinfo {pages} {F339} (\bibinfo {year} {2007})}\BibitemShut
  {NoStop}%
\bibitem [{\citenamefont {Levine} and\ \citenamefont
  {Cockayne}(2008)}]{Levine_2008}%
  \BibitemOpen
  \bibfield  {author} {\bibinfo {author} {\bibfnamefont {Z.~H.}\ \bibnamefont
  {Levine}} and\ \bibinfo {author} {\bibfnamefont {E.}~\bibnamefont
  {Cockayne}},\ }\href {\doibase 10.6028/jres.113.023} {\bibfield  {journal}
  {\bibinfo  {journal} {J. Res. Natl. Inst. Stand. Technol.}\ }\textbf
  {\bibinfo {volume} {113}},\ \bibinfo {pages} {299} (\bibinfo {year}
  {2008})}\BibitemShut {NoStop}%
\bibitem [{\citenamefont {Henkel} and\ \citenamefont
  {Intravaia}(2010)}]{Henkel_2009a}%
  \BibitemOpen
  \bibfield  {author} {\bibinfo {author} {\bibfnamefont {C.}~\bibnamefont
  {Henkel}} and\ \bibinfo {author} {\bibfnamefont {F.}~\bibnamefont
  {Intravaia}},\ }\href {\doibase 10.1142/S0217751X10049608} {\bibfield
  {journal} {\bibinfo  {journal} {Int. J. Mod. Phys. A}\ }\textbf {\bibinfo
  {volume} {25}},\ \bibinfo {pages} {2328} (\bibinfo {year} {2010})},\ \bibinfo
  {note} {special issue ``Quantum field theory in presence of external boundary
  conditions'', edited by Kimball A. Milton and Michael Bordag}\BibitemShut
  {NoStop}%
\bibitem [{\citenamefont {Sernelius}(2005)}]{Sernelius_2005a}%
  \BibitemOpen
  \bibfield  {author} {\bibinfo {author} {\bibfnamefont {B.~E.}\ \bibnamefont
  {Sernelius}},\ }\href@noop {} {\bibfield  {journal} {\bibinfo  {journal}
  {Phys. Rev. B}\ }\textbf {\bibinfo {volume} {71}},\ \bibinfo {pages} {235114}
  (\bibinfo {year} {2005})}\BibitemShut {NoStop}%
\bibitem [{\citenamefont {Svetovoy} and\ \citenamefont
  {Esquivel}(2006)}]{Svetovoy06a}%
  \BibitemOpen
  \bibfield  {author} {\bibinfo {author} {\bibfnamefont {V.~B.}\ \bibnamefont
  {Svetovoy}} and\ \bibinfo {author} {\bibfnamefont {R.}~\bibnamefont
  {Esquivel}},\ }\href {\doibase 10.1088/0305-4470/39/21/S79} {\bibfield
  {journal} {\bibinfo  {journal} {J. Phys. A}\ }\textbf {\bibinfo {volume}
  {39}},\ \bibinfo {pages} {6777} (\bibinfo {year} {2006})}\BibitemShut
  {NoStop}%
\bibitem [{\citenamefont {Brownnutt}\ \emph {et~al.}(2015)\citenamefont
  {Brownnutt}, \citenamefont {Kumph}, \citenamefont {Rabl}, and\ \citenamefont
  {Blatt}}]{Brownnutt_2015}%
  \BibitemOpen
  \bibfield  {author} {\bibinfo {author} {\bibfnamefont {M.}~\bibnamefont
  {Brownnutt}}, \bibinfo {author} {\bibfnamefont {M.}~\bibnamefont {Kumph}},
  \bibinfo {author} {\bibfnamefont {P.}~\bibnamefont {Rabl}},  and\ \bibinfo
  {author} {\bibfnamefont {R.}~\bibnamefont {Blatt}},\ }\href {\doibase
  10.1103/RevModPhys.87.1419} {\bibfield  {journal} {\bibinfo  {journal} {Rev.
  Mod. Phys.}\ }\textbf {\bibinfo {volume} {87}},\ \bibinfo {pages} {1419}
  (\bibinfo {year} {2015})}\BibitemShut {NoStop}%
\bibitem [{\citenamefont {Sipe}(1987)}]{Sipe_1987}%
  \BibitemOpen
  \bibfield  {author} {\bibinfo {author} {\bibfnamefont {J.~E.}\ \bibnamefont
  {Sipe}},\ }\href {\doibase 10.1364/JOSAB.4.000481} {\bibfield  {journal}
  {\bibinfo  {journal} {J. Opt. Soc. Am. B}\ }\textbf {\bibinfo {volume} {4}},\
  \bibinfo {pages} {481} (\bibinfo {year} {1987})}\BibitemShut {NoStop}%
\bibitem [{\citenamefont {Agarwal}(1975)}]{Agarwal_1975a}%
  \BibitemOpen
  \bibfield  {author} {\bibinfo {author} {\bibfnamefont {G.~S.}\ \bibnamefont
  {Agarwal}},\ }\href {\doibase 10.1103/PhysRevA.11.230} {\bibfield  {journal}
  {\bibinfo  {journal} {Phys. Rev. A}\ }\textbf {\bibinfo {volume} {11}},\
  \bibinfo {pages} {230} (\bibinfo {year} {1975})}\BibitemShut {NoStop}%
\bibitem [{\citenamefont {Lifshitz} and\ \citenamefont
  {Pitaevskii}(1980)}]{Landau_vol9}%
  \BibitemOpen
  \bibfield  {author} {\bibinfo {author} {\bibfnamefont {E.~M.}\ \bibnamefont
  {Lifshitz}} and\ \bibinfo {author} {\bibfnamefont {L.~P.}\ \bibnamefont
  {Pitaevskii}},\ }\href@noop {} {\emph {\bibinfo {title} {Statistical Physics
  (Part 2)}}},\ \bibinfo {edition} {2nd}\ ed.,\ \bibinfo {series} {Landau and
  Lifshitz, Course of Theoretical Physics}, Vol.~\bibinfo {volume} {9}\
  (\bibinfo  {publisher} {Pergamon},\ \bibinfo {address} {Oxford},\ \bibinfo
  {year} {1980})\BibitemShut {NoStop}%
\bibitem [{\citenamefont {Buhmann}(2012)}]{Buhmann_vol2}%
  \BibitemOpen
  \bibfield  {author} {\bibinfo {author} {\bibfnamefont {S.~Y.}\ \bibnamefont
  {Buhmann}},\ }\href {\doibase 10.1007/978-3-642-32466-6} {\emph {\bibinfo
  {title} {Dispersion Forces II -- Many-Body Effects, Excited Atoms, Finite
  Temperature and Quantum Friction}}},\ \bibinfo {series} {Springer Tracts in
  Modern Physics}, Vol.\ \bibinfo {volume} {248}\ (\bibinfo  {publisher}
  {Springer},\ \bibinfo {address} {Heidelberg},\ \bibinfo {year}
  {2012})\BibitemShut {NoStop}%
\end{thebibliography}
%

\end{document}